\documentclass[%
reprint,
superscriptaddress,
nofootinbib,                   
 amsmath,amssymb,
 aps,
 pra,
]{revtex4-1}

\usepackage{graphicx}
\usepackage{dcolumn}
\usepackage{bm}
\usepackage[colorlinks,linkcolor=blue,citecolor=blue,urlcolor=blue]{hyperref}
\usepackage{mathtools}
\usepackage{float}
\usepackage{array}
\usepackage{braket}
\usepackage{xcolor}
\usepackage[export]{adjustbox}
\usepackage[normalem]{ulem}          
\usepackage{color}

\begin{document}

\newcommand{\spsc}[1]{\textsuperscript{#1}}
\newcommand{\sbsc}[1]{\textsubscript{#1}}
\newcommand{\mab}[1]{\mathbf{#1}}
\newcommand{\diff}{ \ensuremath{\mathrm{d}} }
\newcommand{\drdm}{ \ensuremath{ \widetilde \rho_{\textrm{1RDM}} } }
\newcommand{\mtex}[1]{ \ensuremath{\textrm{#1}} }
\newcommand{\pmtex}{\ensuremath{\pm}}
\newcommand{\alow}{\ensuremath{\hat{a}}}
\newcommand{\ainc}{\ensuremath{\hat{a}^{\dagger}}}
\newcommand{\mncfg}{\ensuremath{ \textrm{[Ar]} 4s^{2} 3d^{5} (^{6}S_{5/2}) }}
\newcommand{\mnioncfg}{\ensuremath{ \textrm{[Ar]} 4s 3d^{5} (^{7}S_{3}) }}
\newcommand{\mndioncfg}{\ensuremath{ \textrm{[Ar]} 3d^{5} (^{6}S_{5/2}) }}
\newcommand{\mnresnt}{\ensuremath{ 3p^{5} 4s^{2} 3d^{6} (^{6}P) }}
\newcommand{\mnionresnt}{\ensuremath{ 3p^{5} 4s 3d^{6} (^{7}P) }}
\newcommand{\lrarrow}{\ensuremath{ \leftrightarrow }}

\preprint{APS/123-QED}

\title{Time-Dependent Multiconfiguration Self-Consistent-Field Study on Resonantly Enhanced High-Harmonic Generation from Transition Metal Elements}

\author{Imam S. Wahyutama}
\email{iswahyutama@atto.t.u‑tokyo.ac.jp}
\affiliation{
  Department of Nuclear Engineering and Management, Graduate School
  of Engineering, The University of Tokyo, 7-3-1 Hongo, Bunkyo-ku, Tokyo
  113-8656, Japan
}

\author{Takeshi Sato}
\email{sato@atto.t.u‑tokyo.ac.jp}
\affiliation{
  Department of Nuclear Engineering and Management, Graduate School of
  Engineering, The University of Tokyo, 7-3-1 Hongo, Bunkyo-ku, Tokyo 113-8656,
  Japan
}
\affiliation{
  Photon Science Center, Graduate School of Engineering, The University of Tokyo,
  7-3-1 Hongo, Bunkyo-ku, Tokyo 113-8656, Japan
}
\affiliation{
  Research Institute for Photon Science and Laser Technology, The University of
  Tokyo, 7-3-1 Hongo, Bunkyo-ku, Tokyo 113-0033, Japan
}

\author{Kenichi L. Ishikawa}
\email{ishiken@n.t.u‑tokyo.ac.jp}
\affiliation{
  Department of Nuclear Engineering and Management, Graduate School of
  Engineering, The University of Tokyo, 7-3-1 Hongo, Bunkyo-ku, Tokyo 113-8656,
  Japan
}
\affiliation{
  Photon Science Center, Graduate School of Engineering, The University of
  Tokyo, 7-3-1 Hongo, Bunkyo-ku, Tokyo 113-8656, Japan
}
\affiliation{
  Research Institute for Photon Science and Laser Technology, The University of
  Tokyo, 7-3-1 Hongo, Bunkyo-ku, Tokyo 113-0033, Japan
}

\date{\today}

\begin{abstract}
We theoretically study high-harmonic generation (HHG) from transition metal elements Mn and Mn\spsc{+}, using full-dimensional, all-electron, first-principles simulations.
The HHG spectra calculated with the time-dependent complete-active-space self-consistent-field (TD-CASSCF) and occupation-restricted multiple-active-space (TD-ORMAS) methods exhibit a prominent peak at $\sim 50$ eV, successfully reproducing resonant enhancement observed in previous experiments [R.A. Ganeev {\it et al.}, Opt.~Express {\bf 20}, 25239 (2012)].
Artificially freezing $3p$ orbitals in simulations results in its disappearance, which shows the essential role played by $3p$ electrons in the resonant harmonics (RH).
Further transition-resolved analysis unambiguously identifies constructively interfering $3p$-$3d$ ($m=0,\pm 1$) giant resonance transitions as the origin of the RH, as also implied by its position in the spectra.
Time-frequency analysis indicates that the recolliding electron combines with the parent ion to form the upper state of the transitions.
In addition, this study shows that the TD-CASSCF and TD-ORMAS methods can be applied to open-shell atoms with many unpaired inner electrons.

\end{abstract}

\pacs{Valid PACS appear here}
\maketitle


\section{Introduction}\label{sec:introduction}
High intensity and ultrashort laser pulses have become an indispensable tool both in scientific
researches and industrial applications for studying or manipulating the properties of
matter. High-harmonic generation (HHG) is one of the important research domains that have emerged thanks to the remarkable advancement of high-intensity ultrashort laser technologies. HHG is a nonperturbatively nonlinear optical process
in which a fundamental strong-laser field is converted into harmonics of very high orders upon
interaction with atoms, molecules, and solids. 
The nature of the HHG process is closely intertwined with the electronic structure and dynamics of the generating medium. Hence, a variety of quantum scale phenomena have been succesfully identified by devising
specialized measurement techniques such as electronic structure
detection \cite{PhysRevLett.119.203201}, observation of Rabi flopping \cite{PhysRevLett.114.143902}, multi-channel interference \cite{Li_Gui-Hua:224208}, and spectroscopy
of Cooper minimum \cite{PhysRevLett.110.033006,app8071129}.

High-harmonic (HH) radiation is an excellent source of coherent XUV photons that
can fit into a labroom \cite{0034-4885-67-8-C01,RevModPhys.81.163,0034-4885-60-4-001}. A number of applications for studying atomic and material
properties have been reported \cite{PhysRevLett.110.033006,app8071129,PhysRevLett.114.143902,PhysRevLett.119.203201,Li_Gui-Hua:224208} demonstrating its potential as a reliable tool for studying light-matter interaction in the attosecond time scale. 
Further increase in harmonic intensity is desirable to fully explore its areas of use.

It has frequently been reported that the use of transition metal plasma as a generating medium leads to resonant enhancement of a single or a few harmonics \cite{Ganeev:06,Suzuki:06,Ganeev:07,Ganeev:12,
  Ganeev:11,PhysRevA.85.023832,PhysRevA.75.063806,1367-2630-15-1-013051}.
The resonant harmonics (RH) lie close to the giant transition lines in the absorption or emission response of the elements \cite{Ganeev:06}, e.g., $27$ eV in Sn plasma \cite{Suzuki:06,Ganeev:11,PhysRevA.85.023832,fareed2017high}, $20$ eV in In \cite{Ganeev:06}, and 50 eV in Mn \cite{Ganeev:12}. 
The RH generation would offer an attractive way to increase HHG yield around a certain photon energy. It might also serve as a new platform to explore multielectron dynamics in intense laser fields using HHG, which is usually considered to be of single electron nature in most cases. 

In this work, we theoretically investigate the resonant HHG process, using Mn and its cation Mn$^+$ as target systems for the scrutiny of the underlying mechanism. The resonance in HHG spectra from Mn\spsc{+} was observed experimentally by Ganeev \textit{et al.} \cite{Ganeev:12} where the harmonic peak around 50 eV is enhanced by more than an order of magnitude relative to neighboring harmonics, and the 3$p$-3$d$ resonance was suggested to be relevant, as implied by the peak position \cite{Ganeev:12}.

Prior to the present work, there have been theoretical efforts on RH from transition metal elements over
the past decade. Milo{\v s}evi{\' c} in Ref. \cite{0953-4075-40-17-005} and \cite{Milosevic:06}
studied the effect of coherent superposition in the initial state and found that a three-step
process starting from an excited state but returning to the ground one exhibits an enhanced
harmonic. In Ref. \cite{PhysRevA.89.053833}, the line shape of resonant harmonic is discussed
in terms of Fano lineshape. A modelling of the autoionizing state is performed in Refs. \cite{1367-2630-15-1-013051,fareed2017high,PhysRevA.84.013430,PhysRevA.82.023424} using a parametrized potential barrier. Milo{\v s}evi{\' c} has also
investigated the property of resonant harmonic such as intensity and phase in
Ref. \cite{PhysRevA.81.023802}. Other reports of varying elaboration have also been published, see
e. g. Ref. \cite{PhysRevLett.104.123901,PhysRevA.78.053406,PhysRevA.65.023404}. 
Most of the above-mentioned attempts use an effective model potential within the single-active-electron approximation. 
Although not targeted at transition metal plasma, one particular work that starts to consider the possibility of multielectron effects
has been done by Redkin and Ganeev \cite{PhysRevA.81.063825}, who have simulated a fullerene-like model
system using multiconfiguration time-dependent Hartree-Fock (MCTDHF) method within the two-active-electron jelliumlike sphere approximation.

In the present work, we do all-electron three-dimensional (3D) {\it ab initio} simulations based on the time-dependent multiconfiguration self-consistent-field (TD-MCSCF) methods \cite{7115070}, which describe the system wavefunction by the superposition of Slater determinants consisting of time-dependent spin-orbital functions. 
Specifically, we apply state-of-the-art implementation of the time-dependent complete-active-space self-consistent-field (TD-CASSCF) \cite{PhysRevA.88.023402,PhysRevA.94.023405,PhysRevA.97.023423} and time-dependent occupation-restricted multiple-active-space (TD-ORMAS) \cite{PhysRevA.91.023417} methods, which classify spatial orbitals into doubly occupied {\it core} and correlated {\it active} orbitals.
Previously, these methods have been applied to either closed-shell systems or systems having a single unpaired valence electron \cite{PhysRevLett.118.203202, PhysRevA.94.023405, PhysRevA.91.023417, LiYang}. 
Here, we extend our methods to general open-shell atoms such as transition metals, having many unpaired electrons ($5$ and $6$ for Mn and Mn\spsc{+}, respectively) that can equally participate in the dynamics under strong laser fields.
We successfully reproduce the RH at $\sim 50$ eV and unambiguously identify the 3$p$-3$d$ giant resonance as its origin,
by taking full advantage of TD-CASSCF and TD-ORMAS to analyze transition dynamics between different orbitals.
Our results show that the three 3$p$-3$d$ lines ($m=0, \pm 1$) constructively interfere to form the RH peak.

This paper is organized in the following way. 
The overview of the two methods used for our simulations is given in
Sec.~\ref{sec:methods}. The results are presented and discussed in Sec.~
\ref{sec:results}. 
Conclusions and future possibilities are given in Sec.~\ref{sec:conclusion}.
Hartree atomic units are used throughout unless otherwise noted.

\section{TD-CASSCF and TD-ORMAS Methods} \label{sec:methods}

We consider an $N$-electron atom (or ion) with atomic number $Z$ irradiated by a laser field $E(t)$ linearly polarized along the $z$ axis.
In the velocity gauge and within the dipole approximation, its dynamics is described by the time-dependent Schr\"{o}dinger equation,
  \begin{eqnarray}
    \label{eq:SE}
    i\frac{\partial}{\partial t} \Psi(t) &=& \left[ \sum_{i=1}^{N} \left(\frac{-\nabla_{i}^{2}}{2} - \frac{Z}{r_{i}} -iA(t)\frac{\partial}{\partial z_i} \right) \right. \nonumber\\
                          &+& \left. \sum_{i=1}^{N}\sum_{j>i}^{N} \frac{1}{|\mab{r}_{i} - \mab{r}_{j}|} \right] \Psi(t),
  \end{eqnarray}
  with $A(t) = - \int E(t)dt$ being the vector potential.

In the TD-CASSCF and TD-ORMAS methods, we express the total wave function as,
\begin{equation}
	\Psi (t) = \hat{A}\left[\Phi_{\rm fc}\Phi_{\rm dc}(t)\sum_I\Phi_I(t) C_I(t)\right],
\end{equation}
where $\hat{A}$ denotes the antisymmetrization operator, $\Phi_{\rm fc}$ and $\Phi_{\rm dc}$ the closed-shell determinants formed with $n_{\rm fc}$ time-independent doubly occupied frozen-core and time-dependent doubly occupied $n_{\rm dc}$ dynamical-core orbitals, respectively, and $\{\Phi_I\}$ the determinants constructed from  $n_{\rm a}$ active orbitals. 
Whereas in the TD-CASSCF method the active electrons are fully correlated among the active orbitals within prescribed numbers of up- and down-spin electrons, the TD-ORMAS method further subdivides the active orbitals into an arbitrary number of subgroups, specifying the minimum and maximum number of electrons accommodated in each subgroup. 

\begin{figure}
     \begin{center}
        \includegraphics[width=\linewidth]{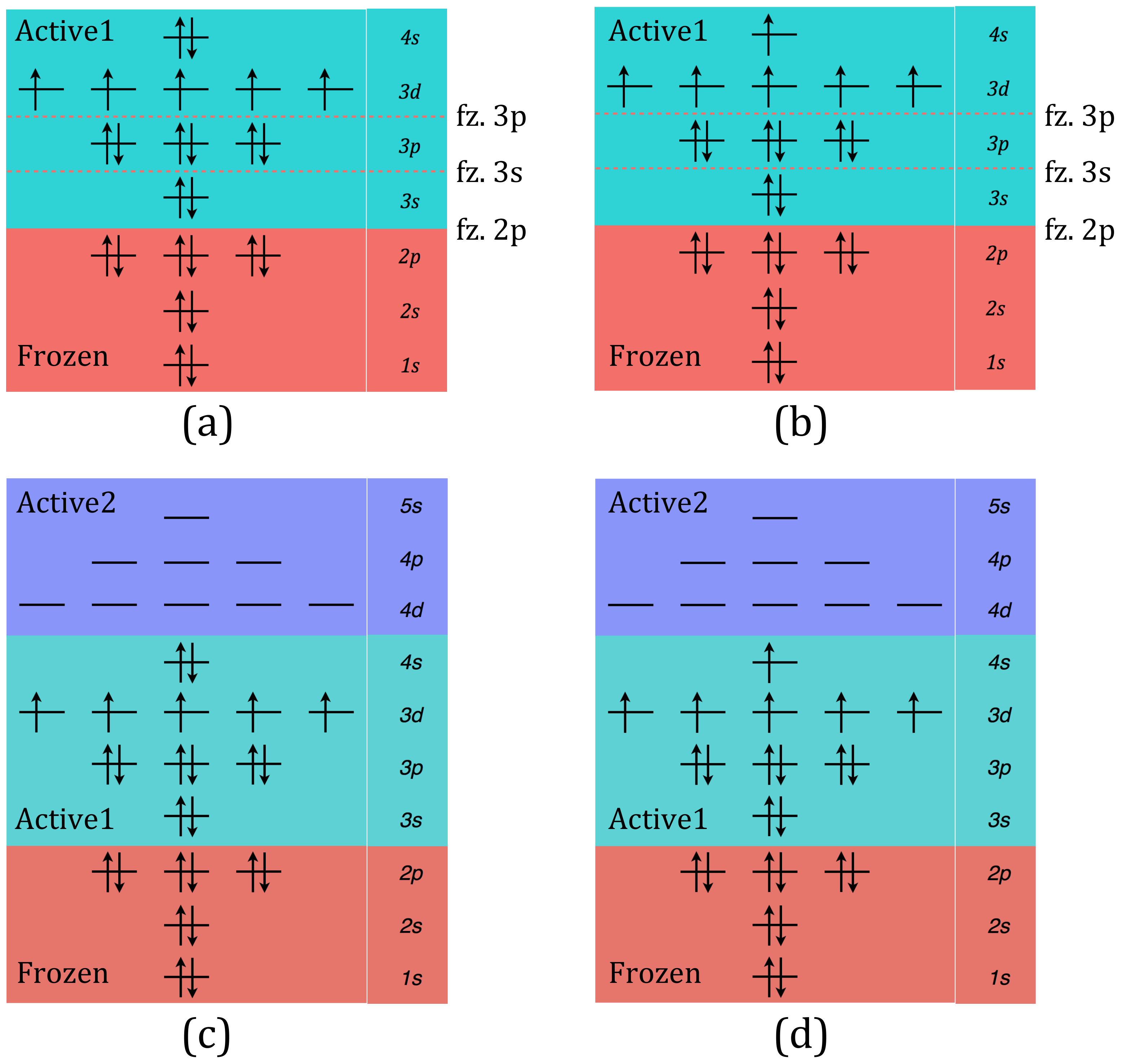}         
     \end{center}
     \caption{\label{fig:orb_dgram}Orbital diagram used for TD-CASSCF with $15$ orbitals for (a) Mn and (b) Mn\spsc{+}. All the electron arrangements within Active1 are allowed. The various frozen spaces are indicated by the dashed lines. For simulation with TD-ORMAS $24$ orbitals, the schemes in (c) and (d) are used for Mn and Mn\spsc{+} respectively, restricting up to two electrons in Active2.}
  \end{figure}

We specifically consider Mn and Mn\spsc{+} in the present study, whose ionization potential, barrier-suppression intensity \cite{1367-2630-15-1-013051,Ilkov_1992}, and the ground-state configuration are summarized in Table \ref{tab:atom_properties}.
Orbital subspace decomposition used in this study is shown in Fig.~\ref{fig:orb_dgram}.
Note that at least $15$ spatial orbitals are required for the correct spin multiplicities. 
TD-CASSCF simulations use $52$ and $44$ determinants for Mn and Mn\spsc{+}, respectively.
In TD-ORMAS simulations, up to two-electron excitations from Active1 to Active2 are allowed, which results in 86510 determinants for Mn and 66068 for Mn\spsc{+}. 

  \begin{table}[]
      \caption{Experimental ionization potential $I_p$, barrier-suppression intensity $I_{BS}$, and the ground-state configuration of Mn, Mn\spsc{+}, and Mn\spsc{2+}.} \label{tab:atom_properties}
      \begin{tabular}{c | c c c }
         \hline \hline
                     & Mn & Mn\spsc{+} & Mn\spsc{2+} \\
         \hline
          $I_p$\footnote{Experimental ionization potential in eV \cite{aip_jpr6_1253}.} & 7.43 & 15.64 & 33.67\\
          $I_{BS}$\footnote{Barrier-suppression intensity in W/cm\spsc{2}.} & $1.2 \times 10^{13}$ & $2.4\times 10^{14}$ & $5.2\times 10^{15}$ \\
          GS\footnote{Ground state configuration \cite{aip_jpr6_1253}.} & \mncfg{} & \mnioncfg{} & \mndioncfg{} \\
         \hline
      \end{tabular}
  \end{table}

The equations of motion (EOMs) describing the temporal evolution of the CI coefficients $\{C_I(t)\}$ and the orbitals $\{\psi_p(t)\}$ are derived on the basis of the time-dependent variational principle (TDVP) \cite{Reinhard1977,LOWDIN19721,Moccia:7.4,Heller:64.1} and read,
\begin{gather}
    \label{eomci}
    i \frac{d}{dt} C_I(t) = \sum_J \braket{\Phi_I| \hat{H} - \hat{R}| \Phi_J}\\
    \label{eomorb}
	i \frac{\partial}{\partial t} \ket{\psi_p} = \hat{h} \ket{\psi_p} + \hat{Q} \hat{F}  \ket{\psi_p} + \sum_{q} \ket{\psi_q}  R^q_p,
\end{gather}
where $\hat{H}$ denotes the total Hamiltonian, $\hat{h}$ the one-body Hamiltonian, $\hat{Q} = 1 - \sum_{q} \Ket{\psi_q} \Bra{\psi_q}$ the projector onto the orthogonal complement of the occupied orbital space.
$\hat{F}$ is a non-local operator describing the contribution from the interelectronic Coulomb interaction, defined as 
\begin{gather}
	\hat{F} \ket{\psi_p} = \sum_{oqsr} (D^{-1})^o_p P^{qs}_{or} \hat{W}^r_s \ket{\psi_q},
\end{gather}
where $D$ and $P$ are the one- and two-electron reduced density matrices, and $\hat{W}^r_s$ is given, in the coordinate space, by
\begin{gather}
	W^{r}_{s} \left({\bf r} \right) = \int d {\bf r}^\prime \frac{\psi_{r}^{*} ({\bf r}^\prime) \psi_{s} ( {\bf r}^\prime )}{| {\bf r} - {\bf r}^\prime | } .
	\label{eq:W}
\end{gather}
The matrix element $R^q_p$ is given by,
\begin{gather}
\label{eq:orbital-time-derivative}
	R^q_p = i \braket{\psi_q | \dot{\psi_p} } - h^q_p,
\end{gather}
with $h^q_p = \braket{\psi_q|\hat{h}|\psi_p}$. 
$R^q_p$'s within one orbital subspace (frozen core, dynamical core and each subdivided active space) can be arbitrary Hermitian matrix elements, and in this paper, they are set to zero. 
On the other hand, the elements between different orbital subspaces are determined by the TDVP. Their concrete expressions are given in Ref.~\cite{PhysRevA.91.023417}, where $iX^q_p = R^q_p + h^q_p$ is used for working variables.
  
Our numerical implementation \cite{PhysRevA.94.023405} employs a spherical harmonics expansion of orbitals with the radial coordinate discretized by a finite-element discrete variable representation \cite{PhysRevA.62.032706,McCurdy_2004,PhysRevE.73.036708,quant_dyn_imag}.
Specifically, to obtain the ground states we use 12 finite elements each of which contains 25 grid points. 
For subsequent time-dependent simulations, up to $62$ finite elements (including those used in the ground state) are employed. 
The initial ground state is obtained through imaginary time propagation of the EOMs.
The Hartree-Fock energies ($-1149.866252$ a.u. for Mn and $-1149.649383$ a.u. for Mn\spsc{+}) perfectly match the values reported in Ref. \cite{aip_jcp103_3000}. 
For the CASSCF and ORMAS cases, electron correlation leads to the lowering of the ground-state energy.
For example, the ORMAS method yields the ground state energy of Mn to be $-1150.0760$ a.u..
We confirm that the total orbital and spin angular momenta of the ground state match the term notations for both Mn ($^6S_{5/2}$) and Mn\spsc{+} ($^7S_3$).

We calculate HHG spectra as the magnitude squared of the Fourier transform of the dipole acceleration $a(t)$, defined as \cite{PhysRevA.94.023405},
  \begin{align}
    \label{eq:accel_define}
    a(t) &=  \sum_{i=1}^{N} \frac{\diff^{2}}{\diff t^{2}} \langle \Psi(t) | z_{i} |
            \Psi(t)\rangle \nonumber \\
         &= -Z\sum_{i=1}^{N} \langle \Psi(t) | \frac{z_{i}}{r_{i}^{3}} | \Psi(t)\rangle - NE(t) + \Delta(\dot{p}_{z}).
  \end{align}
 Here, the additional term $\Delta(\dot{p}_{z})$ accounts for the correction to the Ehrenfest formula in the presence of frozen core orbitals. 
 Its explicit expression is found in Ref.~\cite{PhysRevA.94.023405}.

  \section{Results and Discussions} \label{sec:results} 
  
  \subsection{Resonant high-harmonic emission from Mn and Mn\spsc{+}} \label{sec:act_frz_hhg}
  
  \begin{figure}[tb]
    \includegraphics[width=1\linewidth, right]{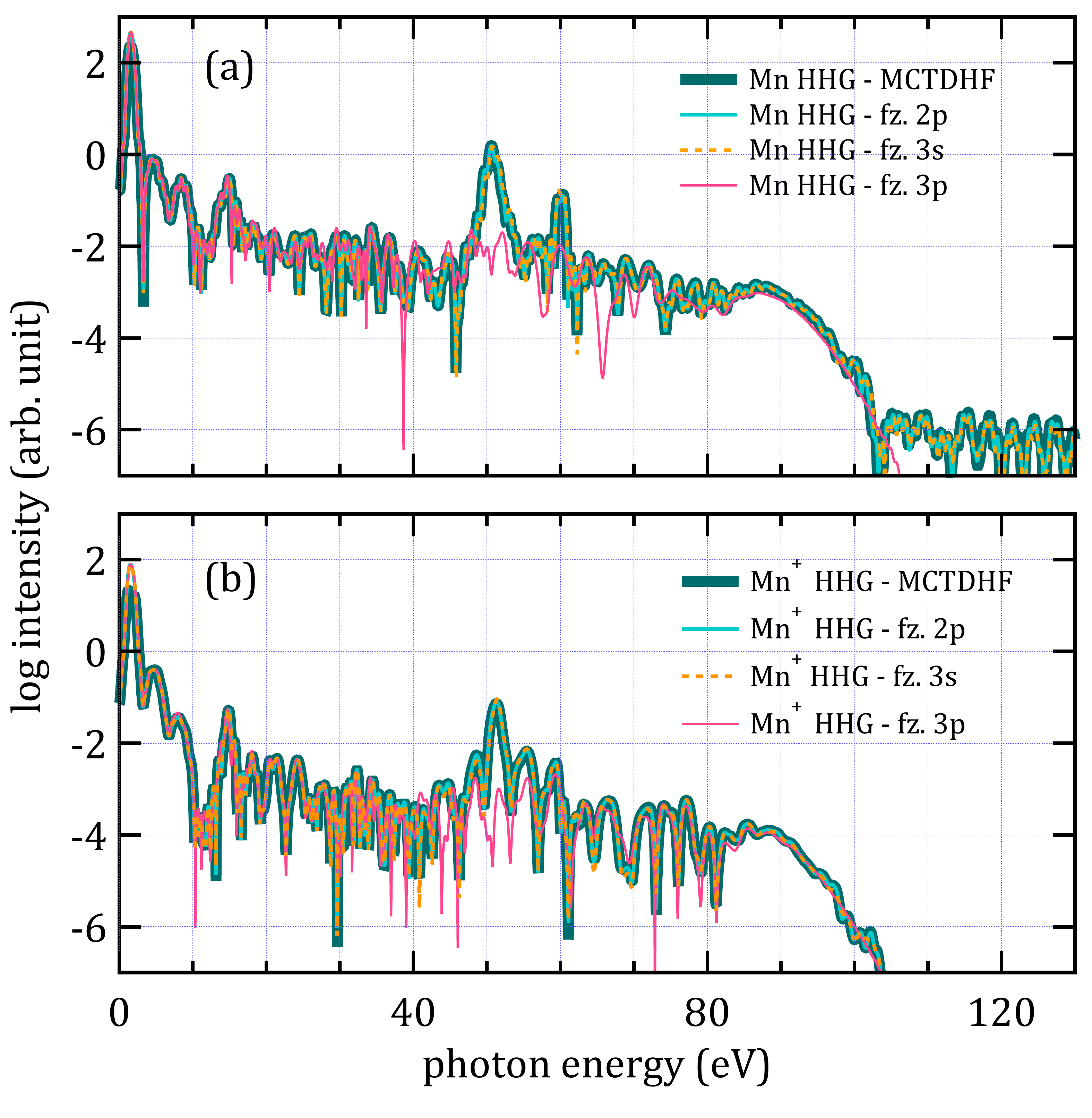}
    \caption{HHG spectra calculated with the TD-CASSCF method for several frozen-core settings shown in Fig. \ref{fig:orb_dgram}(a) and (b) for (a) Mn and (b) Mn\spsc{+}.
    The results of the MCTDHF simulations, in which all the 15 orbitals in Fig. \ref{fig:orb_dgram}(a) and (b) are treated as active, are also shown.
    }\label{fig:HHG_CASSCF_15orbs}
  \end{figure}
  
\begin{figure}[tb]
    \includegraphics[width=1\linewidth, right]{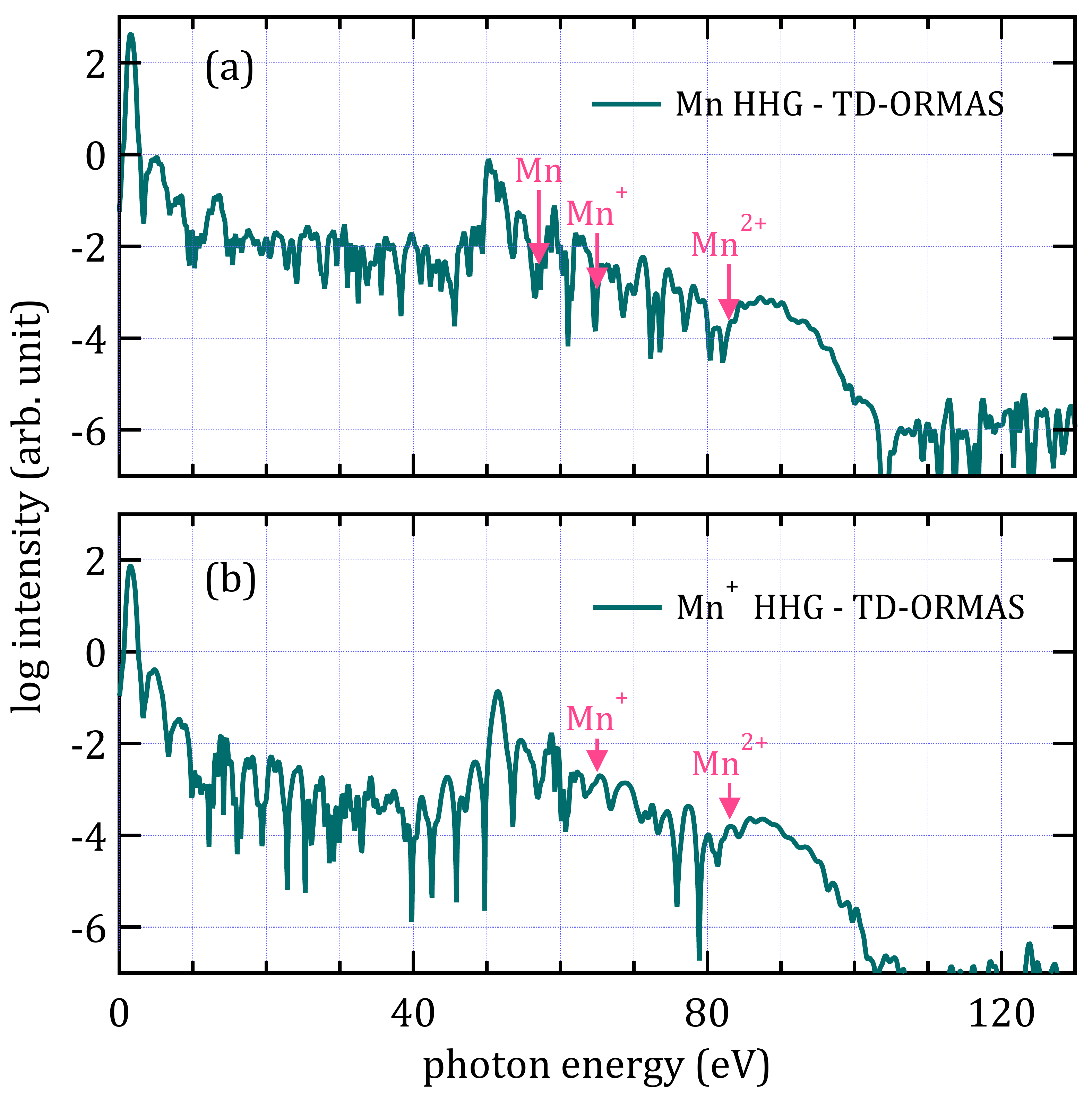}
    \caption{HHG spectra calculated with the TD-ORMAS method for (a) Mn and (b) Mn\spsc{+}. The vertical arrows mark the cutoff positions expected from the three-step model \cite{PhysRevLett.71.1994,super-intense}. \label{fig:HHG_act_vs_frz}}
  \end{figure}

  Let us first examine if our simulations reproduce the resonant harmonics in the HH response of Mn and Mn\spsc{+}. 
  The harmonic spectra from Mn and Mn\spsc{+} obtained with the TD-CASSCF method for a fundamental laser field with $770$ nm central wavelength, $3\times 10^{14}$ W/cm\spsc{2} peak intensity, and foot-to-foot four-cycle $\sin^2$ pulse shape are shown in Fig.~\ref{fig:HHG_CASSCF_15orbs}(a) and (b) (blue solid, marked as ``fz. $2p$"), respectively.
  The results of the MCTDHF simulations, in which all the 15 orbitals in Fig. \ref{fig:orb_dgram}(a) and (b) are treated as active, are also shown (green thick solid curves).
  The perfect overlap of the results by the two methods indicates numerical convergence for this number of orbitals. 
  The results of the TD-ORMAS simulations are plotted in Fig.~\ref{fig:HHG_act_vs_frz} for the same laser parameters as in Fig.~\ref{fig:HHG_CASSCF_15orbs}.
  In both Figs.~\ref{fig:HHG_CASSCF_15orbs} and \ref{fig:HHG_act_vs_frz}, we clearly see an RH slightly above 50 eV, substantially enhanced in comparison with neighboring harmonics, for both Mn and Mn\spsc{+}, whose position is in excellent agreement with the experimental value ($\sim 50$ eV \cite{Ganeev:12}). 

\begin{figure}[tb]
    \includegraphics[width=1\linewidth, right]{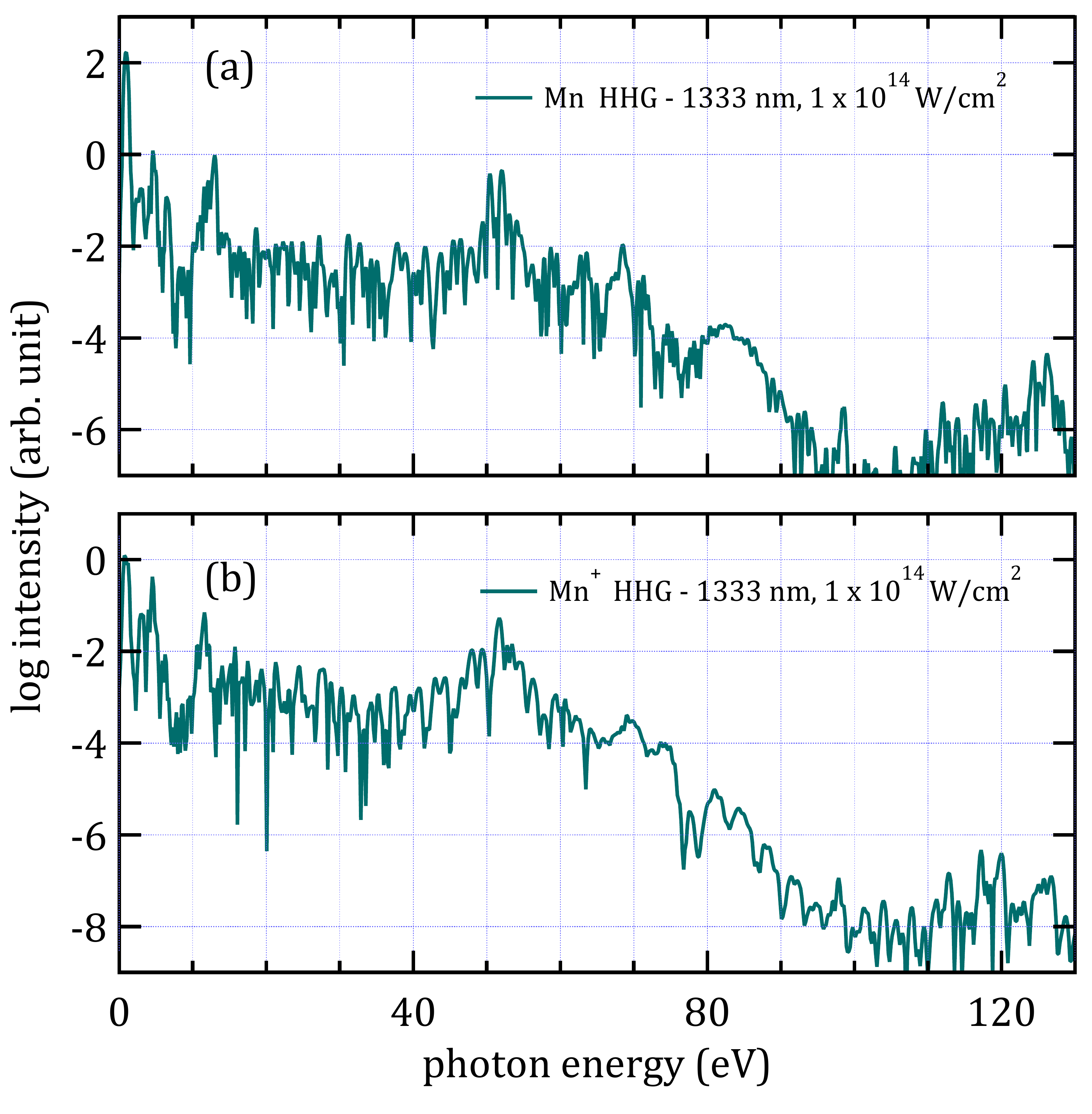}
    \caption{HHG spectra from (a) Mn and (b) Mn\spsc{+} calculated with the TD-ORMAS method for 1333 nm wavelength and $10^{14}\,{\rm W/cm}^2$ peak intensity. \label{fig:HHG_other_lasers}}
  \end{figure}

We have calculated HH spectra for 1333 nm wavelength and $10^{14}\,{\rm W/cm}^2$ peak intensity with the TD-ORMAS method (Fig. \ref{fig:HHG_other_lasers}), while keeping the ponderomotive energy unchanged. In spite of substantial difference in laser parameters, we can see that the resonant harmonic peak remains at $\sim 50$ eV. Thus, the RH position is governed by atomic properties, rather than laser parameters.
  
Then, we make use of the flexibility in orbital-subspace decomposition to find out which orbitals contribute to RH.
We have performed TD-CASSCF simulations by varying the boundary between the active and frozen spaces in Fig.~\ref{fig:orb_dgram}.
Whereas freezing $3s$ virtually does not change the spectrum, freezing up to $3p$ leads to the disappearance of the RH peak (Fig.~\ref{fig:HHG_CASSCF_15orbs}).
This indicates that the appearance of the enhanced peak involves the dynamics of $3p$ electrons. 

 \begin{figure}[tb]
    \includegraphics[width=1\linewidth, right]{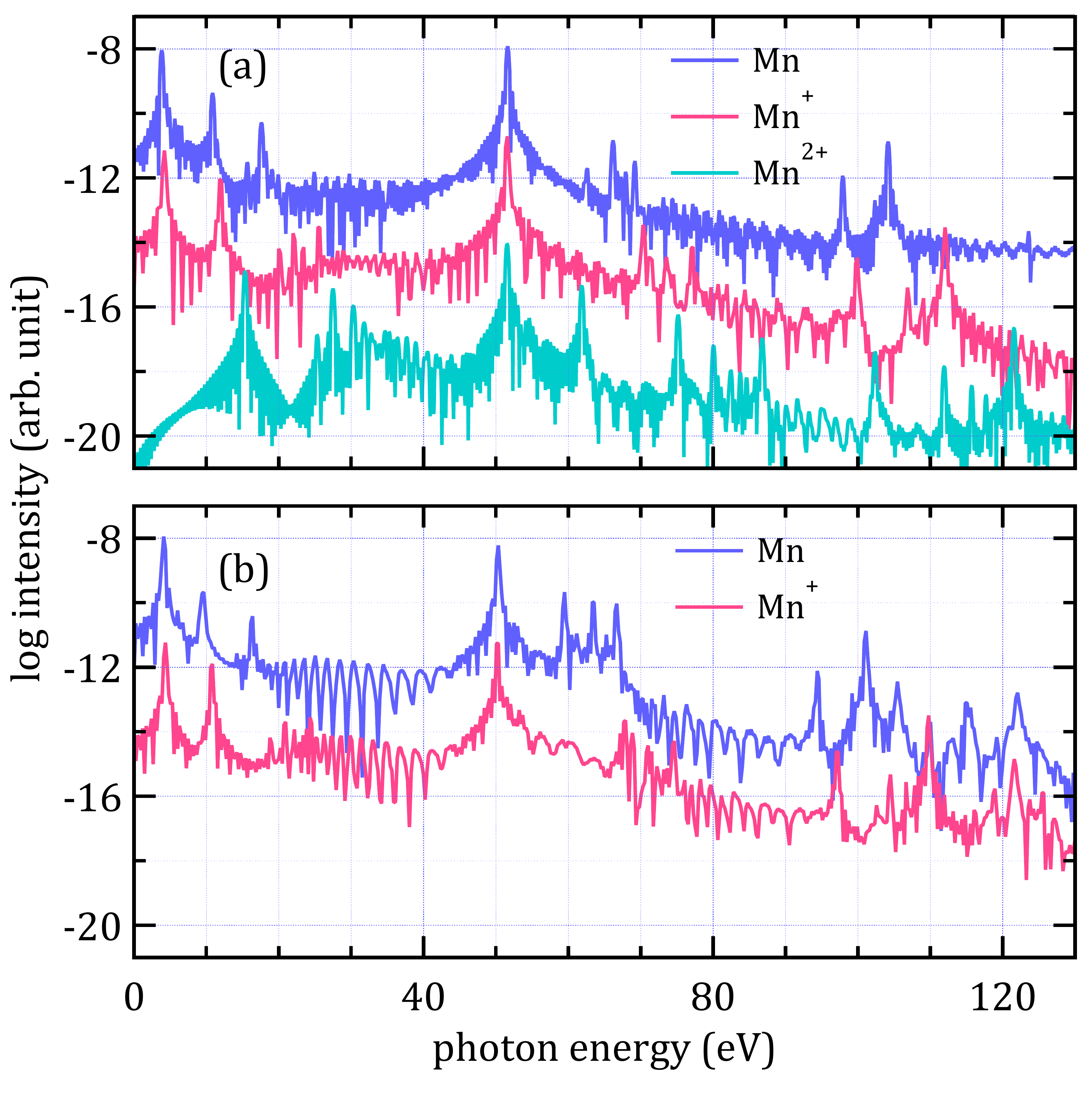}
    \caption{Photoexcitation spectra calculated with (a) TD-CASSCF and (b) TD-ORMAS. The orbital subspace decomposition in Fig. \ref{fig:orb_dgram}(b) without $4s$ orbital is used for Mn\spsc{2+} in (a). The spectra of Mn\spsc{+} and Mn\spsc{2+} are multiplied by $10^{-3}$ and $10^{-6}$, respectively, for better visibility. The excitation spectrum of Mn\spsc{2+} could not be stably calculated with the TD-ORMAS method. \label{fig:linresp}}
  \end{figure}
  
In Fig.~\ref{fig:linresp} are shown the (single-photon) excitation spectra of Mn, Mn\spsc{+}, and Mn\spsc{2+}, obtained as a Fourier transform of the dipole response to a quasi-delta-function pulse with the field being finite at three time steps ($10^{8}$ W/cm\spsc{2} peak intensity).
Although there are slight differences between the TD-CASSCF and TD-ORMAS results, we see a strong excitation line at $\sim 50$ eV in all the cases, which reproduces the position of the well-known $3p$-$3d$ giant resonance line \cite{PhysRevA.39.6074,PhysRevA.62.052703} and coincides with the RH in the HHG spectra. 
These observations strongly suggest that the RH originates from the $3p$-$3d$ resonance line, as also implied experimentally \cite{Ganeev:12}.

 \begin{figure}
    \includegraphics[width=1\linewidth, right]{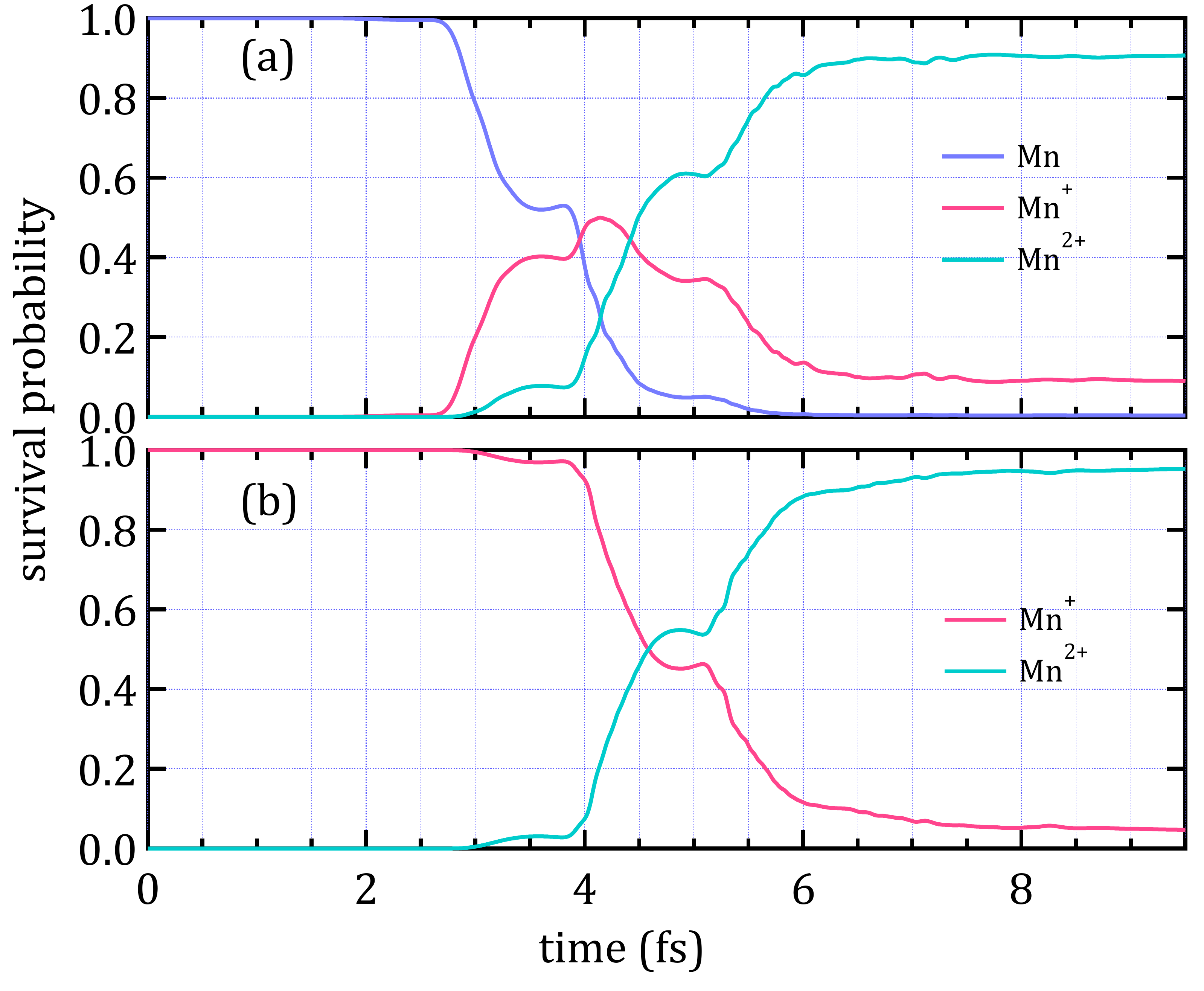}
    \caption{Temporal evolution of the survival probabilities of Mn, Mn\spsc{+}, and Mn\spsc{2+} during the TD-ORMAS simulations starting from (a) Mn and (b) Mn\spsc{+} under the conditions used in Fig.~\ref{fig:HHG_act_vs_frz}. \label{fig:survival_prob}}
  \end{figure}

\begin{figure}
    \includegraphics[width=1\linewidth, right]{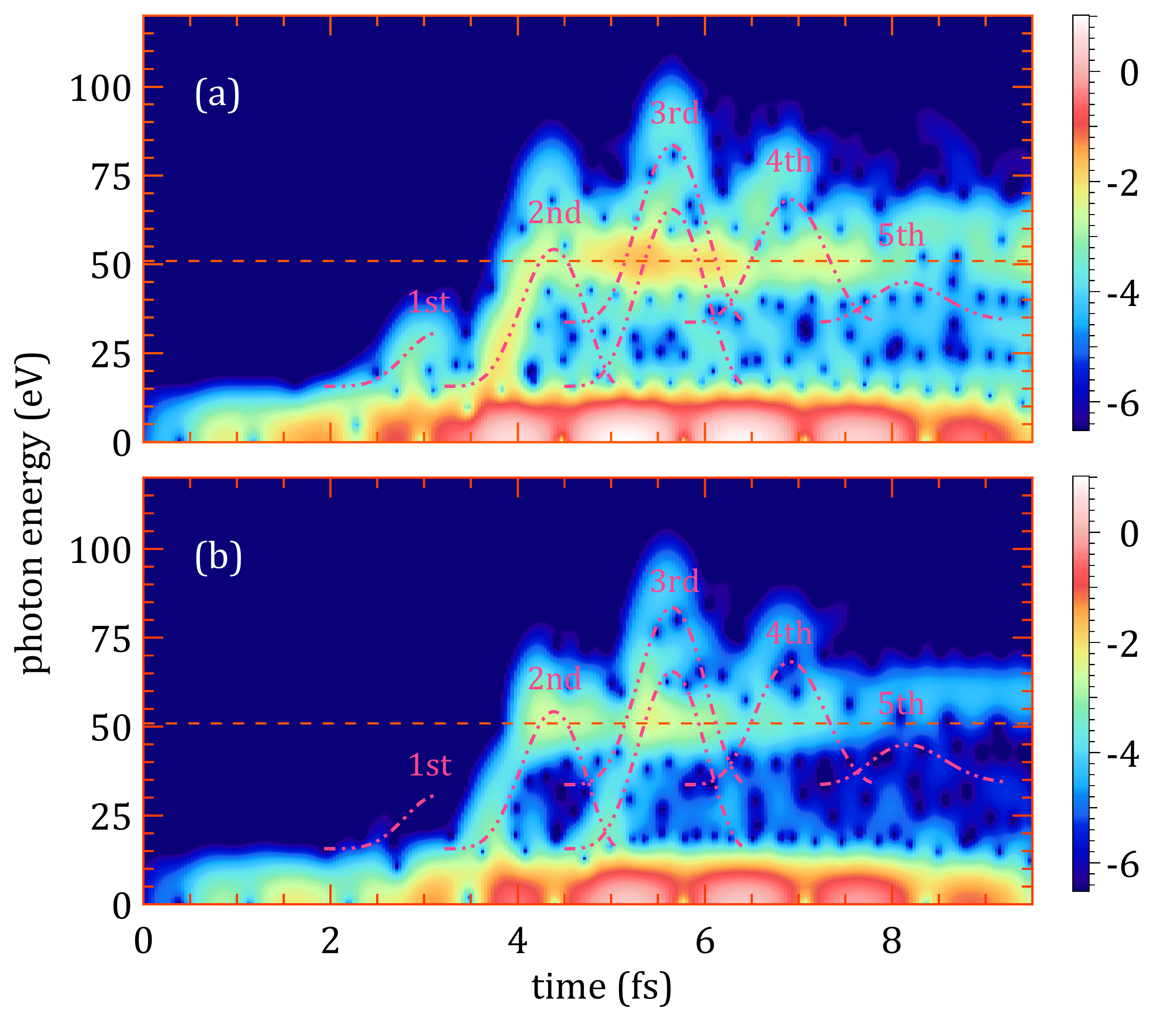}    
    \caption{The time frequency spectrograms of HHG from (a) Mn and (b) Mn\spsc{+} obtained by Gabor transforming the TD-ORMAS results (Fig.~\ref{fig:HHG_act_vs_frz}) with a time window size of 7.5 a.u.. The lower and upper groups of recombination energy (sum of the kinetic energy of the returning electron and the ionization potential) curves are for Mn\spsc{+} and
      Mn\spsc{2+}, respectively.\label{fig:tf_spectro}}
  \end{figure}

Before ending this subsection, let us briefly discuss the cutoff energies.  
The arrows in Fig. \ref{fig:HHG_act_vs_frz} mark the cutoff positions $E_c$ expected from the cutoff law $E_c=I_p + 3.17 U_p$ with $U_p$ being the ponderomotive energy. 
Even if the simulation starts from Mn or Mn\spsc{+}, the HHG spectra extend further up to the cutoff corresponding to Mn\spsc{2+}. 
Indeed, as expected from the barrier-suppression intensity (Table \ref{tab:atom_properties}) and confirmed by Fig.~\ref{fig:survival_prob} showing the temporal variation of the fraction of each species, Mn is mostly ionized in the early stage, and Mn\spsc{+} is further substantially ionized to Mn\spsc{2+}.
The comparison between Fig.~\ref{fig:survival_prob} and the time-freqency structure of HHG shown in Fig.~\ref{fig:tf_spectro} also indicates that the higher plateau appears after the production of Mn\spsc{2+}.
In spite of its high ionization potential, harmonic response of Mn\spsc{2+} is enhanced probably through laser-induced electron recollision \cite{PhysRevLett.118.203202,PhysRevA.97.043414,LiYang}.

\subsection{Transition-resolved analysis} \label{sec:trans_contrib} 
    
    The results in the previous Subsection motivate us to analyze contributions from individual transition lines.
  For the transition-resolved analysis, let us rewrite the dipole acceleration Eq.~\eqref{eq:accel_define} as,
  
  \begin{eqnarray}
    \label{eq:accel_define2}
    a(t) &=& \sum_{m,n} \, \langle m| \hat f |n \rangle \langle n|D(t)|m \rangle - NE(t) + \Delta(\dot{p}_{z})\nonumber\\
         &=& 2\, \sum_{n} \sum_{m>n} \Re\big\{ \langle m| \hat{f} |n \rangle \langle n| D(t) |m \rangle
             \big\}, \nonumber \\
         &-& NE(t) + \Delta(\dot{p}_{z}),         
  \end{eqnarray}
  where $\hat{f} = -Z(z/r^3)$, and $\{|n \rangle\}$ denotes the initial orbitals, obtained through imaginary-time relaxation.
  Since each initial orbital has a definite parity, the terms for $m=n$ vanish. Then, we can view,
  \begin{align} \label{eq:transition}
    \alpha(m,n,t) = 2\, \Re\big\{ \langle m| \hat{f} |n \rangle
    \langle n| D(t) |m \rangle \big\}
  \end{align}
  as a contribution from a transition between orbital pair $m$ and $n$ to the dipole acceleration. 

  \begin{figure}[tb]
    \includegraphics[width=1\linewidth, right]{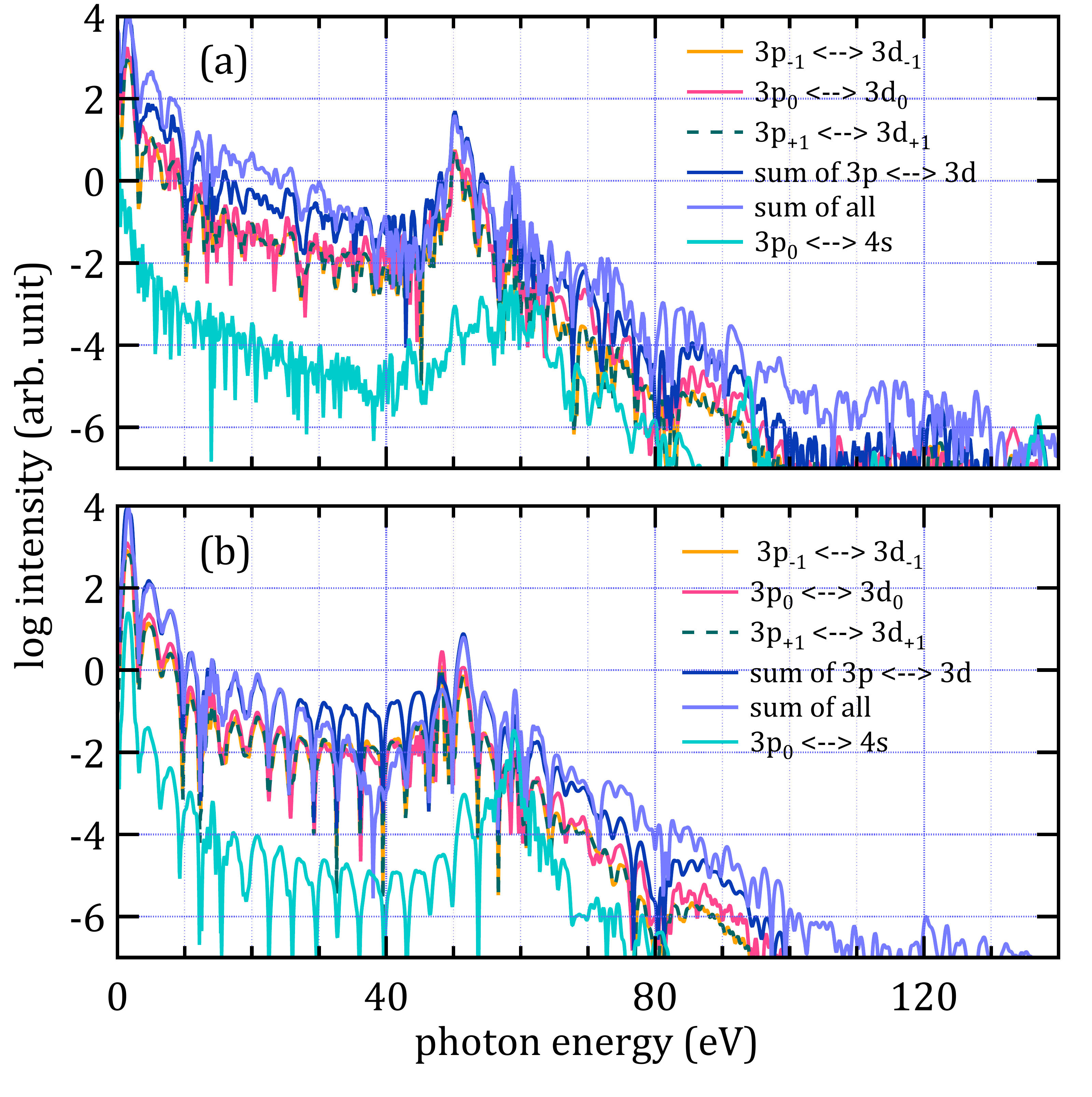}
    \caption{Power spectra of each $3p\leftrightarrow 3d$ transition, $3p\leftrightarrow 4s$, the sum of the three $3p\leftrightarrow 3d$ transition components as well as the sum of all the $18$ transitions listed in Eq.~\eqref{eq:transitions} for (a) Mn and (b) Mn\spsc{+}. \label{fig:trans}}
  \end{figure}
  With the orbitals used in TD-ORMAS simulations [Fig. \ref{fig:orb_dgram}(c) and (d)], we can identify the following $18$ transitions satisfying the selection rule:
  \begin{align}
  \label{eq:transitions}
    & 3s      \lrarrow 3p_{0      } &  & 3s      \lrarrow 4p_{0      } \nonumber\\
    & 3p_{0,\pm 1}\lrarrow 3d_{0,\pm 1} &  & 3p_{0,\pm 1}\lrarrow 4d_{0,\pm 1} \nonumber\\
    & 3p_{0      }\lrarrow 4s       &  & 3p_{0      }\lrarrow 5s       \nonumber\\
    & 4s      \lrarrow 4p_{0      } &  & 3d_{0,\pm 1}\lrarrow 4p_{0,\pm 1} \nonumber\\
    & 4p_{0,\pm 1}\lrarrow 4d_{0,\pm 1} &  & 4p_{0      }\lrarrow 5s      ,
  \end{align}
  where the subscripts denote the magnetic quantum numbers. 
  We have calculated the power spectrum of each transition line as the magnitude squared of Fourier transform of $\alpha(m,n,t)$.

The spectrum of each of $3p_{0,\pm 1}\lrarrow 3d_{0,\pm 1}$, the contribution of their sum, and the total spectrum of all the lines listed in Eq.\eqref{eq:transitions} are presented in Fig. \ref{fig:trans} for Mn and Mn\spsc{+}.
The sum contribution of $3p_{0,\pm 1}\lrarrow 3d_{0,\pm 1}$ has a clear peak at $\sim 50$ eV and dominates the total contribution in that photon energy region. 
The other transitions involving $3p$ such as $3p_0\leftrightarrow 4s$ are by orders of magnitude weaker.
Moreover, the position and form of the peak agree very well with those of the RH in the HHG spectra shown in Fig.~\ref{fig:HHG_act_vs_frz}.
It should also be noticed that the sum spectrum of the three $3p_{0,\pm 1}\lrarrow 3d_{0,\pm 1}$ lines is approximately one order of magnitude stronger than the contribution of each transition.
These observations unambiguously establish that the resonant harmonic at $\sim 50$ eV, experimentally discovered \cite{Ganeev:12} and numerically reproduced by our {\it ab initio} simulations is driven by a constructive interference of electron dynamics occurring between $3p$ and $3d$ orbitals.

Superposed on the spectrograms in Fig.~\ref{fig:tf_spectro}, we plot, as a function of time, the recombination energy, defined as the sum of the kinetic energy of the returning electron and the ionization potential, or, harmonic photon energy expected from the three-step model \cite{PhysRevLett.71.1994,super-intense}.
We see that RH photons are emitted mainly when the recombination energy is $\sim 50$ eV, where the recolliding electron has to be recaptured by the parent ion in order to induce a $3p \to 3d$ transition.
This is consistent with recombination to autoionizing states (the upper states of the $3p$-$3d$ giant resonance lines in the present case), proposed in Refs.~\cite{1367-2630-15-1-013051,fareed2017high,PhysRevA.84.013430,PhysRevA.82.023424}.
Whereas these studies used the single-active-electron approximation with a model potential barrier, our all-electron {\it ab initio} simulations support this process as a mechanism of RH generation from Mn and Mn\spsc{+}.

   \section{Conclusions} \label{sec:conclusion}

We have applied the TD-CASSCF and TD-ORMAS methods to open-shell elements to study resonant enhancement in high-harmonic generation from transition metal elements Mn and Mn\spsc{+}.
Our simulations have successfully reproduced the presence and position ($\sim 50$ eV) on the experimentally observed resonance harmonics \cite{Ganeev:12}.
While its position suggests the relevance with the $3p$-$3d$ giant resonance lines, we have performed a series of analyses to unambiguously verify it.
First, we have taken advantage of flexibility in orbital subspace decomposition to vary the boundary between frozen-core and active orbitals, and found that freezing up to $3p$ leads to the disappearance of the RH, which shows the essential role of the $3p$ electrons.
Then, we have calculated the contribution of each transition between initial orbitals to harmonic spectra.
It has indeed revealed that the RH is dominated by the $3p$-$3d$ ($m=0,\pm 1$) transitions, constructively interfering.
It has followed from the inspection of the HHG time-frequency structure that the RH is emitted mainly, if not exclusively, when the sum of the kinetic energy of the returning electron and the ionization potential of Mn or Mn\spsc{+} is $\sim 50$ eV.
This implies that the electron recombines to the autoionizing upper state of the $3p$-$3d$ transitions, as proposed previously within the single-active-electron approximation using a model potential barrier \cite{1367-2630-15-1-013051,fareed2017high,PhysRevA.84.013430,PhysRevA.82.023424}.

\begin{acknowledgments}
This research was supported in part by a Grant-in-Aid for Scientific Research 
(Grants No.~16H03881, No.~17K05070, and No.~18H03891)
from the Ministry of Education, Culture, Sports, Science and Technology (MEXT) of Japan and also 
by the Photon Frontier Network Program of MEXT.
This research was also partially supported by the Center of Innovation Program from the Japan Science 
and Technology Agency (JST), by CREST (Grant No.~JPMJCR15N1), JST, and by Quantum Leap Flagship Program of MEXT.
I.S.W. gratefully acknowledges support from Special Graduate Program in Resilience Engineering of the University of Tokyo.
\end{acknowledgments}

   \bibliography{%
     0034-4885-67-8-C01,
     RevModPhys.81.163,
     0034-4885-60-4-001,
     PhysRevLett.110.033006,
     app8071129,
     PhysRevLett.114.143902,
     PhysRevLett.119.203201,
     Li_Gui-Hua-224208,
     PhysRevLett.100.173001,
     PhysRevLett.103.073902,
     PhysRevLett.98.013901,
     PhysRevLett.99.253903,
     PhysRevA.49.2117,
     PhysRevA.80.011807,
     jap113.6.063102,
     0953-4075-40-17-005,
     Milosevic-06,
     PhysRevA.89.053833,
     PhysRevA.84.013430,
     PhysRevA.82.023424,
     PhysRevA.81.023802,
     PhysRevA.81.063825,
     MEYER199073,
     BECK20001,
     PhysRevLett.104.123901,
     PhysRevA.78.053406,
     PhysRevA.65.023404,
     PhysRevA.39.6074,
     PhysRevA.62.052703,
     0953-4075-31-5-008,
     0953-4075-37-6-014,
     PhysRevA.43.1441,
     PhysRevA.54.974,
     aip_jpr6_1253,
     PhysRevA.88.023402,
     7115070,
     PhysRevA.94.023405,
     PhysRevA.91.023417,
     PhysRevA.97.023423,
     PhysRevA.81.053845,
     PhysRevLett.118.203202,
     Ganeev-06,
     Suzuki-06,
     Ganeev-07,
     Ganeev-12,
     Ganeev-11,
     PhysRevA.85.023832,
     PhysRevA.75.063806,
     1367-2630-15-1-013051,
     fareed2017high,
     PhysRevA.60.3125,
     PhysRevA.81.043408,
     Reinhard1977,
     LOWDIN19721,
     Moccia-7.4,
     Heller-64.1,
     PhysRevA.62.032706,
     McCurdy_2004,
     aip_jcp103_3000,
     Ilkov_1992,%
     PhysRevE.73.036708,%
     quant_dyn_imag,%
     PhysRevA.97.043414,%
     PhysRevLett.71.1994,%
     super-intense,%
     LiYang%
   }

\end{document}